\documentclass[useAMS,usenatbib]{mn2e}
\usepackage{graphicx}

\def\lta{\lower2pt\hbox{$\buildrel {\scriptstyle <} 
   \over {\scriptstyle\sim}$}}
\def\gta{\lower2pt\hbox{$\buildrel {\scriptstyle >} 
   \over {\scriptstyle\sim}$}}

\begin{document}

\title[Unified GRB and X-ray afterglow emissions]{A unified picture for gamma-ray burst prompt and X-ray 
afterglow emissions}
\author[Kumar et al.]{P. Kumar,$^1$ E. McMahon$^1$,
  S. D. Barthelmy$^2$,  D. Burrows$^3$, 
  N. Gehrels$^2$, M. Goad$^4$,
  \newauthor J. Nousek$^3$, and G. Tagliaferri$^5$ \\
$^1$Department of Astronomy, University of Texas, Austin, TX 78712\\
$^2$NASA, Goddard Space Flight Center, Greenbelt, Maryland 20771 \\
$^3$Department of Astronomy and Astrophysics, 525 Davey Lab,
Pennsylvania State University, University Park, PA 16802 \\
$^4$Department of Physics and Astronomy, University of Leicester,
Leicester LE 1 7RH, UK \\
$^5$INAF-Osservatorio Astronomico di Brera, Via Bianchi
46, I-23807 Merate (LC), Italy}
\maketitle
\begin{abstract}
Data from the Swift satellite has enabled us for the first time
to provide a complete picture of the gamma-ray burst emission
mechanism and its relationship with the early afterglow emissions. We
show that gamma-ray photons for two bursts, 050126 \& 050219A, for which 
we have carried out detailed analysis were produced as a result of 
the synchrotron self-Compton process in the material ejected in the 
explosion when it was heated to a mildly relativistic temperature at 
a distance from the center of explosion of order the deceleration radius. 
Both of these bursts exhibit rapidly declining early X-ray afterglow 
lightcurves; this emission is from the same source that produced the gamma-ray
burst.  The technique we exploit to determine this is very general and
makes no assumption about any particular model for gamma-ray
generation except that the basic radiation mechanism is some
combination of synchrotron and inverse-Compton processes in a relativistic
outflow.  For GRB 050219A we can rule out the possibility that energy 
from the explosion is carried outward by magnetic fields, and that the 
dissipation of this field produced the $\gamma$-ray burst.
\end{abstract}

\section{Introduction}
The successful launch of the Swift satellite in November 2004 filled a
crucial gap in the gamma-ray burst data at early times -- between a
minute to a few hours -- that existed in prior GRB missions. This has
led to a number of very interesting discoveries regarding emission
from GRBs on time scales of minutes following a burst.  One of these
discoveries is that the very early X-ray lightcurve (LC) of many
bursts falls off very rapidly: $f_x \propto t^{-3}$ 
(Tagliaferri et al. 2005, Goad et al. 2005, 
Burrows et al. 2005, Chincarini et al. 2005). This
phase of rapid fall off lasts for about 5 minutes, and is followed by
the usual $f_x \propto t^{-1}$ behavior. In most cases, no change to 
the spectral slope is seen accompanying the change to the lightcurve. In 
this paper, we discuss two bursts exhibiting such behavior, GRBs 050126 
and 050219A. We provide an argument that the $\gamma$-rays and the early
X-rays (for the first $\sim$5 minutes) have a common source, and we
determine the physical properties of the source (next section). 

\section{Modeling Prompt $\gamma$-ray and Afterglow Emissions}

We start with some
very general physical considerations and describe a model with as few
assumptions as possible to try to understand the $\gamma$-ray and
X-ray emissions together.



We do not assume that $\gamma$-rays are produced in the internal
shock or external shock or any other of a number of different models
that have been suggested.  We determine the properties of the
$\gamma$-ray source from the data and use it to decide which of the
proposed models, if any, work.  The only assumption that we make is that
$\gamma$-rays are generated by synchrotron or inverse-Compton (IC)
mechanisms -- an assumption that is supported by their non-thermal
spectrum and also indirectly by the excellent overall agreement
between models based on synchrotron \& IC emission and multiwavelength
afterglow data for a large number of GRBs (Piran, 2005; M\'esz\'aros \& 
Rees, 1999; Panaitescu \& Kumar, 2002; Granot et al. 1999).

The two bursts considered here have $\gamma$-ray light curves dominated 
by a single peak and small fluctuations and therefore much of the
$\gamma$-ray flux is likely produced in a single source localized in
space. In such a case the synchrotron and IC emissions from the object
are completely determined if we know the magnetic field strength
($B$), the optical thickness of the object to Thomson scattering
($\tau_e$), the speed of the object toward the observer
(Lorentz factor -- $\Gamma$), the total number of radiating particles 
assuming isotropic source ($N$), and the
minimum energy for radiating particles, $\gamma_i m_e c^2$, ($m_e$ is
electron mass \& $c$ is the speed of light).  The particle energy
distribution above $\gamma_i$, at the acceleration region where
particles have not suffered appreciable loss of energy, is assumed to
be a power-law function with index $p$.  The energy distribution of
particles for the entire population, however, is not a single power
law function due to the loss of energy via radiative processes. We
determine this modified distribution numerically by carrying out a
self-consistent calculation of synchrotron cooling and self-absorption
frequencies as described in Panaitescu \& M\'esz\'aros (2000), and
McMahon et al. (2005).

The average energy per particle, at the acceleration site, in the
comoving frame of the source is $\epsilon = \gamma_i m_e c^2
(p-1)/(p-2)$, and therefore $N\approx E_{iso}/(\epsilon\Gamma)$; where
$E_{iso}$ is the isotropic equivalent of energy in $\gamma$-rays.
The index $p$ is determined by
the observed spectral index; we take $p=2.4$ when the spectrum above
the peak is not known -- results reported here have been checked for
dependence on $p$, and found qualitatively to be insensitive to $p$.

So we are left with four unknown parameters viz., $B$, $\tau_e$,
$\Gamma$ and $\gamma_i$. The observational constraints on these
parameters are: the $\gamma$-ray flux at the peak of the observed
light curve, the frequency at which
the spectrum peaks, the duration of the burst, the spectral index
below the peak, and the optical flux limit (when available).  The last
two constraints are not independent and typically provide a limit on
synchrotron cooling and/or injection frequencies.

The optical depth, $\tau_e$, and $N$ determine the distance of the
$\gamma$-ray source from the center of the explosion:
$r=\sqrt{N\sigma_T/4\pi\tau_e}$; and the burst duration
$t_{GRB}\approx r/2\Gamma^2c$. The parameters we use describe the
state of the $\gamma$-ray producing source at the time of the 
peak of the observed lightcurve. The observed peak flux is the
synchrotron or inverse-Compton flux in the appropriate observer
energy band which is determined from $B$, $N$ (the total number of 
electrons/positrons in the source), $\tau_e$, $\gamma_i$ and $\Gamma$;
the details of the calculation is described in Kumar et al. (2005). 
By searching the parameter space
($B$, $\tau_e$, $\Gamma$, $\gamma_i$) for emission properties
consistent with those observed for each burst we can decide among
various GRB models.  As we shall see, we are led to more or less a
unique solution: $\gamma$-rays are generated via synchrotron-self-IC (SSC)
in a source with typical electron energy less than 10$^3m_e c^2$ and
with properties that favor the external reverse-shock
or internal shocks. Moreover, the $\gamma$-ray source we thus find also
accounts for the early X-ray afterglow in a natural way, as off axis
flux from the $\gamma$-ray emitting material or flux from the adiabatically
cooling source.

Results for 050126 and 050219A are discussed below.

\subsection{GRB 050126}

GRB 050126 was 25s long with a fluence in 15-350 kev band 
of $1.7\pm0.3\times10^{-6}$ erg cm$^{-2}$, and 
redshift 1.29 (Tagliaferri et al. 2005). The average spectral index 
$\beta$ ($f_{\nu} \propto \nu^{\beta}$) during the burst was 
$-0.34 \pm 0.14$ and during
the X-ray afterglow, the spectral index was $\beta = -1.35\pm 0.3$ and
the LC fell off as $t^{-2.52^{+0.5}_{-0.2}}$. 
We describe the results for $\gamma$-ray and X-ray emissions below.

\noindent{\it Gamma-ray generation via the synchrotron process}

Figure 1 shows the parameter space allowed -- for a source radiating
via the synchrotron process -- to explain the observed $\gamma$-ray
data for 050126. In particular we show the allowed range for
$\gamma_i$, $B$, $\Gamma_1$ (the lower limit to the Lorentz factor of unshocked
shell which produced the $\gamma$-ray photons when it was shock
heated -- we will refer to it as shell 1), and the upper limit to the
Lorentz factor of the shell or the medium that shell 1 collided 
with ($\Gamma_2$);
the figure caption describes how $\Gamma_1$ and $\Gamma_2$ are
calculated.  These quantities are plotted against the radius, $r$, at
which $\gamma$-rays are generated, to determine which GRB model could
be described by the four parameter solution space.

The solutions we find have $\gamma_i>3000$, and a high magnetic field
strength is needed to explain the $\gamma$-ray
emission for this burst if it were to arise due to synchrotron
emission.  The synchrotron cooling frequency is found to be less than
a few ev which is in part due to the constraint that the low energy
spectral index is -0.34$\pm0.14$ (so all of the solutions are in
highly radiative cooling regime).  
The radius where the observed $\gamma$-rays could have been generated
varies from the typical internal shock radius of $\sim10^{14}$ cm to
the external shock radius of $\sim10^{16}$cm; the lower limit to the
radius is due to our choice of $\tau_e<0.1$ in order to avoid excessive 
Compton scattering --- for $\tau_e=1$ the
minimum $r$ is a factor 2 smaller.  In the case of internal shocks we find
that the Lorentz factor of the two colliding shells must satisfy the condition
$\Gamma_1\gta10^3$ and $\Gamma_2\lta3$ (see fig. 1), which seems an
unrealistic requirement for any central engine to meet, and in any
case this situation would not be that different from the interaction
of GRB ejecta with the ISM where $\Gamma_2=1$. Note that the time
interval between the ejection of the two shells (with $\Gamma_1=10^3$
and $\Gamma_2=2$) is larger than 500s for the internal shock radius of
$r\sim10^{14}$cm while the duration of this burst was 25s -- this is another
problem for this solution. Furthermore, the fact that the GRB LC was a
FRED (fast rise, exponential decline) means that internal shocks are
not required to generate the $\gamma$-ray emission.

The allowed parameter space contains an external forward-shock
solution as well ($r\sim10^{16}$cm; $\Gamma_2=1$). This solution,
however, requires $\Gamma > 10^4$ (figure 1) which makes the 
already acute problem of baryonic loading much worse. Moreover, 
the deceleration radius for this large $\Gamma$, for a typical
GRB-circumstellar medium density of $\sim 10$ cm$^{-3}$, is less 
than $10^{16}$cm -- the distance at which the $\gamma$-ray source        
according to our solution is located. 
Therefore, we conclude that $\gamma$-rays from 050126 are unlikely to
have been produced via the synchrotron process in internal or external
shocks.

\begin{figure}
\includegraphics[width=0.49\textwidth,angle=0]{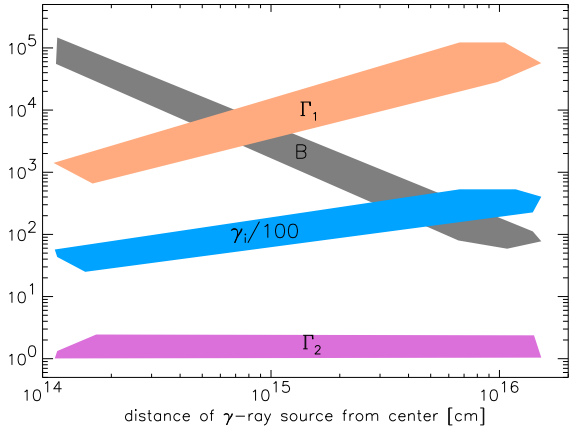}
\caption{The parameter space for the synchrotron
radiation solution to GRB050126.  The x-axis is the radial distance of
the $\gamma$-ray source from the center of explosion.  Shown in the
figure is the minimum energy of electrons divided by the rest mass
($\gamma_i$) at the location where these particles are accelerated in
the source, i.e., where radiative losses are unimportant.  Also shown
are the comoving magnetic field ($B$) in Gauss, the Lorentz factor (LF) of the 
unshocked shells/medium -- $\Gamma_1$ \& $\Gamma_2$.
$\Gamma_1=\Gamma\Gamma_{sh}(1+v v_{sh})$ is the LF of the inner
unshocked shell; where $\Gamma_{sh}=(p-1)\gamma_i m_e/(p-2)m_p$ is the
minimum LF of the shock front wrt the unshocked shell -- this assumes
that electrons have the same energy as protons ($\Gamma_{sh}$ will be
larger if electrons have lower energy) -- and $\Gamma$ is the LF of
the shocked material as seen by a lab frame observer. $\Gamma_2$ is
the LF of the unshocked outer shell/medium and is given by:
$\Gamma\Gamma_{sh}(1-v v_{sh})$.  The calculations of $\Gamma_1$ and
$\Gamma_2$ are valid when the $\gamma$-ray producing shell/medium is
the inner and the outer shell respectively, and they are also valid for
most internal shell collision situations where the shock front speed in
the two shells is about the same. For these calculations we took
$E_{iso}=10^{52}$ erg, $p=2.4$, $z=1.29$, and the flux at 150keV at
the peak of the $\gamma$-ray LC (7s) to be 0.2mJy.  We use a factor of
2 tolerance in all of the observational data such as $\gamma$-ray
flux, burst duration etc. in constructing the acceptable solution
parameter space.}
\end{figure}

\subsubsection{Gamma-ray production via the inverse-Compton process}

Figure 2 shows the allowed parameter space for synchrotron-self-IC solutions.
The entire solution space consists of mildly relativistic shocks with
$2<\gamma_i<1000$, and the Lorentz factor (LF) of the source is between 20 \&
300. Mildly relativistic shocks arise naturally in internal collisions
(with the ratio of LFs for colliding shells of order a few) and the
external reverse-shock (RS).  A good fraction of the allowed parameter
space has electron cooling time, due to radiative losses, of order the 
dynamical time or less,
and the synchrotron cooling frequency is of order a few eV. The magnetic field
strength is about 50 Gauss (which corresponds to $\epsilon_B\sim0.1$) 
\& Compton Y $\sim 1$
for the part of parameter space corresponding to reverse-shocks,
whereas $B$ is between 1 and 10$^3$ gauss and $1\lta Y\lta10^4$ for
internal shocks.  The IC $\gamma$-ray lightcurve falls off very
rapidly for both the internal and the reverse shock emission as does
the observed LC (Kobayashi et al. 2005).  Therefore, $\gamma$-rays from 
GRB050126 could have been produced via SSC in either internal or external
shocks, and we don't see any reason to prefer one solution over the
other for this burst.

\subsubsection{X-ray afterglow}

Is it possible that the early X-ray afterglow was produced by the same
source as the GRB IC photons? The IC cooling frequency,
$\nu_{c}^{IC}\sim \nu_{c} \gamma_c^2$, at the GRB LC peak (7s) is
typically of order a few hundred keV for the allowed parameter space
for this burst. Since $\nu_c^{IC}$ shifts to lower energies due to
adiabatic cooling, as $\sim t^{-2}$, at 100s it will have dropped to
$\sim$1 keV.  In this case the flux in the XRT band at 100s from
$\theta\lta\Gamma^{-1}$ part of the source will be very small, and
will rapidly drop to zero on a short time scale. The early X-ray LC
could be explained by this adiabatically cooling $\gamma$-ray source
provided that $\nu_{c}^{IC}\gta10$ MeV at 7 s, which is somewhat
outside of the parameter space we find for this burst.

Could photons detected by the XRT in 0.2-10 keV band at $t>100$s be
off-axis photons (Kumar \& Panaitescu, 2000) that originate at the source 
at an angle w.r.t. the line of sight $>\Gamma^{-1}$?  The flux at 
10 keV at the peak of the
GRB 050126 lightcurve was 0.54$\pm0.08$mJy. This gives the flux
\footnote{The off-axis flux falls off as $t^{-2+\beta}$, see [11], 
where $\beta$ is the spectral index, i.e. $f_\nu\propto\nu^{-\beta}$.}
at 100s due to off axis emission of 1.1$\pm.15\mu$Jy, in rough agreement
with the XRT measurement of 2.8$\pm1.2\mu$Jy.  The X-ray
light-curve between 100s and 425s declined as
$t^{-2.52^{+0.5}_{-0.22}}$.  This decline is also consistent with that
expected of off-axis emission; $\beta=1.26\pm0.22$ during this period
would give rise to off-axis LC decaying as $t^{-3.26\pm0.22}$.
 The spectral
peak for the off-axis emission from a uniform jet decreases with time
as $1/t$, and so the peak at 100s is at $\sim 10$keV.  The peak frequency
decreases more rapidly when electron energy and/or magnetic field is
smaller at higher $\theta$. In this case the spectral peak will be
below 10 keV, and $\beta$ in the XRT band, for $t>100$s, smaller than
during the GRB. We note that a decrease of $\gamma_i$ and $B$ would
not lead to a decrease in the flux in the XRT band so long as these
changes are accompanied with an increase in the number of radiating
particles as might be expected, for instance, when $\Gamma$ decreases
with $\theta$ but the energy per unit solid angle is roughly constant.
The angular structure of the ejecta can be constrained by the
difference between the observed spectral peak at 100s and during the
burst.

\begin{figure}
\includegraphics[width=0.49\textwidth,angle=0]{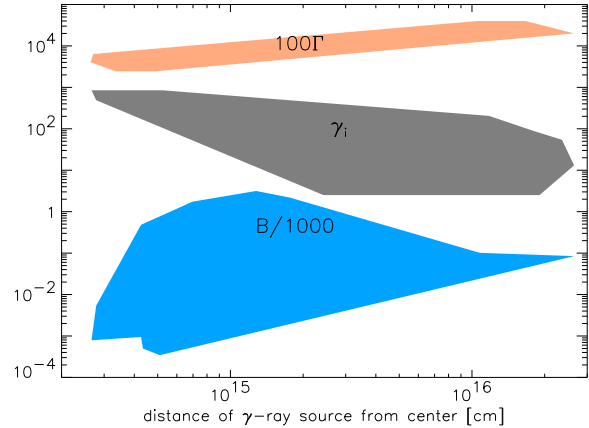}
\caption{The parameter space for synchrotron-self-inverse-Compton solution
to GRB050126. Shown in the figure are allowed range for $\gamma_i$,
$\Gamma$ (the Lorentz factor of $\gamma$-ray source), and $B$ (the comoving 
magnetic field strength in Gauss). See figure 1 caption for some relevant 
details about the calculation.}
\end{figure}

\begin{figure}
\includegraphics[width=0.49\textwidth,angle=0]{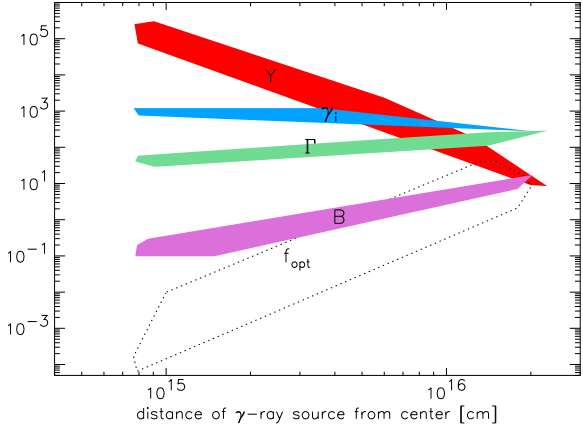}
\caption{ The parameter space for the synchrotron-self-inverse-Compton solution
to GRB050219A. Shown in the figure are allowed range of $\gamma_i$,
$\Gamma$ (the Lorentz factor of $\gamma$-ray source), Compton $Y$ parameter, 
$B$ (the comoving magnetic field strength in Gauss), and the predicted 
optical flux at 100s for these solutions (assuming burst redshift
of 1 and no extinction).
The solutions with $r< 4\times10^{15}$cm have $Y\gta10^4$
and are physically unacceptable since the energy in the 2nd Compton
scattering will be of order 10$^{54}$erg which is too large to obtain
from a stellar mass object. Therefore, the only viable solution for the
$\gamma$-ray emission is IC in the external reverse-shock.
We took $E_{iso}=10^{53}$ erg, $p=2.9$, $z=1$, the peak of the
spectrum at 90keV, and the flux at
the peak of the $\gamma$-ray LC (15s) at 90keV to be 1.2mJy.
We applied the condition that $\nu_a^{IC}\equiv \nu_a\times\min(\gamma_i,
\gamma_c)^2\sim$90 keV; this automatically ensures that $\beta=0.75\pm0.3$
as observed. We use a factor of 2 tolerance in all of the observational
data such as $\gamma$-ray flux, burst duration, the peak frequency
etc. in constructing the acceptable solution parameter space.}
\end{figure}

\subsection{GRB 050219A}

GRB 050219A was 23.6s long with fluence in the BAT 15-350 kev band
of $5.2\pm0.4\times10^{-6}$
erg cm$^{-2}$. The average spectral index $\beta$ during the burst was
$0.75 \pm 0.30$ ($f_\nu\propto\nu^{0.75 \pm 0.30}$), and the peak of 
the spectrum was at 90$\pm9$ keV (Tagliaferri et al. 2005).
During the X-ray afterglow, the spectral index was $\beta=-1.1\pm0.2$
and the X-ray LC declined as $t^{-3.15\pm0.22}$.
We describe below the mechanism for $\gamma$-ray, X-ray, and optical emissions.

\subsubsection{Gamma-ray production}

The positive $\beta$ during the GRB, although consistent with the
synchrotron spectrum of $\nu^{1/3}$ to within 1.5$\sigma$, rules out
the synchrotron process for the generation of $\gamma$-rays for
050219A. The reason is that the magnetic field required to produce
synchrotron peak frequency of 90keV is sufficiently strong that
electrons lose their energy on a time scale much less than the
duration of the burst (23s), and in this case the spectrum below 90
keV would be $\sim\nu^{-1/2}$.
\footnote{A synchrotron frequency of 90 keV implies that
$B\gamma_i^2\Gamma= 10^{13}$ and the electron cooling LF is
$\gamma_c/\gamma_i \sim 10^{-17} \gamma_i^3\Gamma/t_{GRB}(1+Y)$; the
Compton parameter $Y\sim\tau_e\gamma_i\gamma_c$, and therefore,
$(\gamma_c/\gamma_i)^2\sim 10^{-17}\gamma_i\Gamma/(\tau_e t_{GRB})$,
where $t_{GRB}$ is the burst duration in the host galaxy rest frame. Since
$\tau_e>10^{-8}$and $t_{GRB}\sim 10$s, and $\gamma_i<10^3 \Gamma$, we
see that $\gamma_c/\gamma_i<1$ unless $\Gamma>3000$ which is highly
unlikely.}  This is in conflict with the observed spectrum and rules
out the synchrotron process for $\gamma$-ray generation.

The inverse-Compton process on the other hand provides a very natural
way of explaining the observed spectrum and other properties. The
spectrum produced by inverse Compton scattering of a self-absorbed
synchrotron radiation is $f_\nu\propto\nu$ for $\nu< \nu_a\times
\min(\gamma_i, \gamma_c)^2\equiv \nu_a^{IC}$; where $\nu_a$ is the
synchrotron self-absorption frequency. For $\min(\gamma_i,
\gamma_c)\sim 300$ and $\nu_a\sim 1$eV, the peak of the IC spectrum at
$\nu_a^{IC}$ is close to the observed value of 90$\pm$9 keV. These
parameters arise naturally in an external reverse-shock.

Figure 3 shows the allowed parameter space for SSC
solution for GRB 050219A, assuming $z=1$.  The range of $\gamma_i$ for
the allowed solutions is 200--1500 which is typical for the external
reverse-shock and for internal shocks, but not the external
forward-shock. The magnetic field $B$ is between 0.1 and 20
Gauss. This is highly sub-equipartition ($\epsilon_B\lta10^{-3}$), and
therefore for 050219A we can rule out the possibility that the
$\gamma$-ray burst was produced as a result of dissipation of magnetic
field or that much of the energy of the explosion was carried outward
by the magnetic field.  

The Compton Y is rather large -- of order 10--100 for external shock
($r\sim 10^{16}$cm), and larger than 10$^4$ for internal shock radius
of $r\sim 10^{14}--10^{15}$cm (fig. 3). One might suspect that the 
large $Y$ renders 
these solutions unphysical since the energy in the 2nd Compton scattering,
which produces $>$100 GeV photons, will far exceed the $\gamma$-ray
energy.  However, for low optical depth systems with
$\Gamma\gg1$ the radius of the system increases by about a factor 2 in
the time it takes photons to traverse the shell. Therefore, the
optical depth for the 2nd scattering is smaller than the 1st by a
factor 4, and the electron thermal energy has decreased due to
adiabatic expansion during this period by a factor of about 4 for RS
(shell thickness for RS increases as $r^{7/2}$), and a factor 2 for
internal shocks. Thus, the effective $Y$ for the 2nd Compton
scattering is smaller than the 1st scattering Compton-Y by a factor of
about 64 for the RS and 16 for internal shocks. For this reason
$Y\sim100$ for the external shock is quite acceptable, as the total energy
requirement is of order 10$^{51}$ erg.  However,
$Y>10^4$ for internal shocks (see fig. 3) would require the total
energy in the explosion to be $\sim10^3$ times larger than energy in
the $\gamma$-ray band and that is highly unlikely considering that
$E_\gamma\sim10^{51}$ erg.  Therefore, the only viable solution for
the $\gamma$-ray production for 050219A is inverse-Compton in the
external reverse-shock heated ejecta.

\subsubsection{X-ray afterglow}

There are two mechanisms that can explain the X-ray observations for
this burst. One of these is the off-axis emission.  The flux at the
peak of $\gamma$-ray LC (15s) at 10keV was $\sim 300\mu$Jy.  Using
this and $\nu^{0.75\pm0.3}$, we find the flux at 100s, due to the
off-axis emission mechanism ($f_{\nu}\propto t^{-2 + \beta = -1.25 \pm
0.3}$), to be $\sim 29\pm7\mu$Jy, which is consistent with the
observed XRT flux ($25\pm9 \mu$Jy at 10 keV at 100s).  The LC decay
according to the off-axis emission after the spectral peak falls
through the XRT band is $t^{-2+\beta}$, where $\beta=-1.1\pm0.2$ is
the spectral index for $t>100$s, and this is consistent with the
observed decay of $t^{-3.15\pm0.22}$.  The difference between the
X-ray afterglow and $\gamma$-ray spectra can be understood in the same
way as discussed for 050126, i.e. the peak of $f_{\nu}$ during the GRB
(90keV) is well below 10keV at 100s if $\gamma_i$ and $B$ decrease
with $\theta$ slightly and this changes the spectrum from
$\sim\nu^{0.7}$ to $\sim\nu^{-1}$.\footnote{Angular variation is
almost unavoidable, because in the absence of it the early X-ray LC
would have declined as $t^{-1.25}$ due to the off axis emission.}

The second possibility is that we continue to see radiation from
within $\Gamma^{-1}$ angle of the adiabatically cooling $\gamma$-ray
source.  We find that for a large part of the allowed parameter space
for $\gamma$-ray solution $\nu_c^{IC}\gta1$MeV, and therefore we expect
to receive emission in the 0.2-10 keV band for a period of about 5
minutes, during which time the flux decline will be $\sim t^{-2.8}$,
which is consistent with the observed decay.
\footnote{The IC frequencies for an adiabatically cooling ejecta
shift with time as $t^{-2}$, so the 90 keV peak at 15s would have
shifted to 2 keV at 100s. During the time when this peak is above the
XRT band of 10keV, the IC flux from the RS decreases very weakly with
time ($\sim t^{-0.4}$), and subsequently, the flux decreases as
$t^{-2.8}$.  The cross-over is expected at about 45s. Thus, the flux
from the RS at 100s at 10 keV is expected to be about $18\pm4\mu$Jy
which is consistent with the XRT flux of 25$\pm9\mu$Jy. The spectrum
at 100s will be as expected of IC above $\nu_a^{IC}$, i.e. roughly
$\nu^{-1}$. }  We note that the discontinuity in the BAT and XRT
lightcurves for this burst$^1$ could be due to an underestimation of the
spectral evolution in 20-50s time interval where the BAT signal was
low.  A discontinuous jump can also arise in the off-axis model as a
result of a rapid increase in jet energy for $\theta$ between
$\gamma^{-1}$ and $2\gamma^{-1}$.

Both of these solutions suggest a common source
for the $\gamma$-ray burst and early X-rays.

\subsubsection{Optical observations}

The optical flux at 100s from the $\gamma$-ray source is shown in
figure 3.  For the RS solution the flux is about 1 mJy whereas the
observed UVOT upper limit at 96s is 0.02mJy (Schady et al. 2005). The 
much smaller optical flux could be due to absorption in the host
galaxy. The total hydrogen column density for this burst was$^1$ 
$2.2\pm0.6\times10^{21}$ cm$^{-2}$, in excess of the
galactic value, which for a burst at $z\sim1$ could give $\sim$7mag of
optical extinction, more than sufficient to bring the optical flux
below the observed upper limit.\footnote{Galactic correlation between 
$N_H$ and extinction might not
apply to GRBs due to possible dust destruction by GRB emission (Galama
\& Wijers, 2001). It
is therefore difficult to say with confidence the amount of extinction
for this burst in the V-band. } Alternatively, if the RS occurs at
$r\lta 10^{16}$ cm the optical flux would be roughly consistent with
the observed upper limit (see fig. 3).  However, in this case
$Y\sim300$, and the energy in the 2nd Compton scattered, photons, at
100 GeV, will be almost an order of magnitude larger than the energy
in $\gamma$-rays.

\section{Conclusion}

We find that the prompt $\gamma$-ray and early (first few
minutes) X-ray emissions for GRBs 050126 and 050219A are
consistent with being produced by the same source. In the case of
050126, the emission is inverse Compton radiation from either internal
shocks or external-reverse shock, and in the case of 050219A, the
photons are produced by inverse Compton in the external-reverse shock.
The late time X-ray ($t\gta5$min) is produced, as usual, in the 
forward shock.

These results can be applied to the class of gamma-ray bursts with consist
of a simple, i.e. not highly variable, lightcurve. For instance,
our conclusion that $\gamma$-rays were generated via the inverse-Compton 
process for GRB 050219A is valid for all those GRBs which, like GRB 050219A,
have a positive low energy spectral index for the
prompt gamma-ray emission ($f_\nu\propto\nu^{\beta}$ with $\beta>1$).
The allowed values for parameters -- $B$, $\Gamma$ \& $\gamma_i$ --
for the source of $\gamma$-rays for any GRB consisting of a single peak
in the $\gamma$-ray lightcurve should be similar to that shown in 
figures 1 and 3 for 050126 \& 050219A.

{\bf Acknowledgment: } PK thanks Alin Panaitescu and Jonathan Granot for 
useful discussions.  This work is supported in part by grants from NASA 
and NSF (AST-0406878) to PK.

\end{document}